\documentclass[aps,pre,reprint,superscriptaddress, nofootinbib ]{revtex4-2}
\usepackage[utf8]{inputenc}
\usepackage[T1]{fontenc}
\usepackage{amsmath}
\usepackage{amssymb}
\usepackage{amsfonts}
\usepackage{graphicx}
\usepackage{tensor}
\usepackage{xcolor}
\usepackage{cleveref}
\newcommand{\J}[1]{{\color{violet}#1}}

\begin{document}

\title{Computational quantum field theory for fermion pair creation in 2-dimensional curved spacetimes}
\author{Mohammed Alkhateeb}
\affiliation{Research Unit Lasers and Spectroscopies (UR-LLS), naXys \& NISM, University of Namur, Rue de Bruxelles 61, B-5000 Namur, Belgium.}
\affiliation{Mathematical Sciences Department, University of Plymouth,
Plymouth, PL4 8AA, UK}
\author{James P. Edwards}
\affiliation{Mathematical Sciences Department, University of Plymouth,
Plymouth, PL4 8AA, UK}
\author{Yves Caudano}
\affiliation{Research Unit Lasers and Spectroscopies (UR-LLS), naXys \& NISM, University of Namur, Rue de Bruxelles 61, B-5000 Namur, Belgium.}

\begin{abstract}
Similarly to the well-known particle / anti-particle pair production process in strong electromagnetic fields (the Schwinger effect), the matter field vacuum state can be excited by time-dependent, curved spacetime geometries. We study this process using a spacetime resolved numerical approach in the interaction picture by extending Computational Quantum Field Theory (CQFT), well-adapted to simulate the time evolution of quantum fields, to spin-1/2 fermions in curved spacetime. 

Within this framework, we investigate vacuum excitation of a Dirac field induced by a spacetime-curvature quench. In particular, we evolve the fermionic Minkowski vacuum in a 1+1-dimensional idealized curved spacetime characterized by a localized Gaussian deformation of flat spacetime. Particle production is quantified by fermion–antifermion pair numbers defined with respect to the Minkowski basis of asymptotically flat spacetime. We analyze how the excitation depends on the strength and spatial extent of the curvature deformation and discuss the numerical implementation of CQFT in curved backgrounds. While the post-quench geometry is static and no electromagnetic field is included, this work establishes a foundation for studying particle creation in genuinely time-dependent curved spacetimes and electromagnetic backgrounds.
\end{abstract}

\maketitle
\newpage
\section{Introduction}

The creation of particle-antiparticle pairs from the quantum vacuum by strong electromagnetic fields was studied from a field theory perspective by Schwinger \cite{schwinger}, employing the  quantum electrodynamics of fermions in an electric field background. Schwinger built upon earlier work of Sauter, who solved the Dirac
equation in a classical electric field background \cite{sauter}, alluding
to the possibility of particle-antiparticle pair creation. In
particular, Sauter identified transitions between positive- and
negative-energy states induced by potential ramps of the order of the
rest energy over a Compton wavelength. An elegant route to the pair creation rate, via the optical theorem, was provided by the calculation of the imaginary part of the effective action by Euler and Heisenberg (spinor QED) \cite{EHL} and Weisskopf (scalar QED) \cite{WK}.
The Schwinger effect can be understood as a tunneling process out of a false vacuum, and so is exponentially suppressed below a "critical" electric field strength, 
$1.3 \times 10^{18} \textrm{V/m}$. Such fields still cannot be reached in the lab frame in earth-based experiments. Note, however, that the fields experienced in the rest frame of a Lorentz-boosted electron beam incident on a high-intensity laser can surpass the critical field strength, inducing other non-linear processes (for examples of laser facilities see \cite{taya,yoon,luxetec,facetii}).

A related effect was later predicted by Hawking \cite{hawking}, where strong gravitational fields near the event horizon of black holes lead to the creation of particle-antiparticle pairs, giving rise to what is known as Hawking radiation. There, the curved spacetime background plays a role analogous to the electric field in the Schwinger effect. Similarly, the Unruh effect can be associated with the appearance of a horizon in the spacetime of a uniformly accelerated observer \cite{PhysRevD.7.2850, Davies_1975, PhysRevD.14.870}.
The success of quantum field theory in curved spacetimes in predicting Hawking radiation has inspired further studies in the emerging field of gravity analogues, including optical \cite{opt1,opt2} and acoustic \cite{acou1,acou2} versions of black holes, as well as black hole analogues in condensed matter systems \cite{analog_graphene, photonic_chip}.

In the context of strong electromagnetic fields, computational quantum field theory (CQFT) has proven successful in studying pair creation resulting from colliding laser pulses where the created particle densities are obtained by evolving the fermionic vacuum state in the presence of a time-dependent background field \cite{grobe_intro,grobe_arb,grobe_enhance,grobe_phase}. CQFT has also shed light on the dynamics of Klein tunneling \cite{ours_fstq_klein_1,ours_fstq_klein_2,ours_qft_klein} and provided insights into the tunneling time problem \cite{ours_strictly_local,ours_tun_comp}. The extension of CQFT to curved spacetime seems natural given this success, and is the subject of this work. Doing so would align the CQFT framework with recent studies of pair creation in combined electromagnetic and gravitational fields \cite{PhysRevD.98.045015, Chen:2012zn, Frob:2014zka, PhysRevD.52.3742, Garriga:1994bm}. 

Particle creation by time-dependent and spatially varying
gravitational backgrounds has been extensively studied within the
framework of quantum field theory in curved spacetime
\cite{birrell_davies, parker_toms}. Well-known examples include
cosmological particle production during inflation
\cite{parker_original, ford_inflation} and Hawking radiation from black
holes \cite{hawking, fredenhagen1989}. Standard analytic techniques include Bogoliubov transformations
relating asymptotic in- and out-mode  \cite{birrell_davies,parker_original,parker_toms,hawking,ford,ford_inflation,ford_creation_1978} and adiabatic
renormalization schemes \cite{parker_fulling_adascheme}. Alternatives, based on saddle points \cite{Semren:2025dix, Ilderton:2025umd, Akhmedov:2024axn} in the worldline formalism of quantum field theory \cite{UsRep, ChrisRev} and the double copy \cite{ Carrasco:2026ijt, ilderton2025hawking, Carrasco:2025bgu, Aoude:2025jvt} have also been applied. These approaches provide deep conceptual insight and some have been applied to a variety of geometries \cite{Carrasco:2026ijt, ilderton2025hawking, Carrasco:2025bgu, Aoude:2025jvt}. However, their ability to account for real-time field dynamics in regions of spatially localized and time dependent curvature profiles of the type considered here is less straightforward.

On the numerical side, several methods have been developed to solve the
Dirac equation in curved geometries. For studies of Dirac materials, quasiconformal coordinate transformations
have been used to simulate electron dynamics on strained graphene
surfaces, which induces an effective curvature
\cite{fillionPRE2021}. Pseudospectral and operator-splitting
techniques can treat Dirac Hamiltonians with
spatially varying coefficients with high accuracy \cite{fillionpseudo}. These approaches focus on the
single-particle Dirac equation, whereas we focus here on the fully dynamical Dirac field and investigate pair creation induced directly
by spacetime curvature within the CQFT framework. 

The present work contributes to this landscape by extending 
computational quantum field theory (CQFT) 
\cite{grobe_intro, grobe_arb, grobe_phase} to curved spacetimes.
CQFT evolves the fermionic field state in real time by recasting this evolution as generated by the first-quantized Dirac Hamiltonian. This is simulated on a discretized spacetime lattice by constructing the time evolution operator using the split-operator approach to avoid the doubling problem corresponding to spurious fermion modes caused by lattice discretization \cite{fillion_ps2012}.

One challenge in CQFT is the ambiguity of
particle number at transient times (see, for example \cite{ildertonadia}). The expectation value of the particle number (or density) operator, or the contribution to this from given momentum / spin states, is in general dependent on the basis of one-particle states used to define these (second-quantized) operators. This reflects the inequivalence of the asymptotic vacua (and indeed that of the instantaneous Hamiltonian) which lead to different mode decompositions of the field -- which also shows pair creation to be down to vacuum decay. For example, when particle
numbers are computed using a mode expansion in terms of adiabatic eigenstates of the Dirac equation built from the vacuum state in the asymptotic future, the resulting particle numbers correspond to those that would be observed asymptotically if the external background were removed instantaneously \cite{ildertonadia}; these can vary over orders of magnitude larger than the physical, asymptotic number of created pairs. To mitigate this ambiguity,
calculations are typically performed in fields that
vanish asymptotically in space or time, where an unambiguous particle interpretation is available far from the interaction region
\cite{ours_qft_klein,ours_strictly_local,ours_tun_comp}.  
Similarly, in the present work, the extended CQFT framework for curved spacetimes
is applied to backgrounds that are asymptotically flat.
Importantly, however, a central advantage of CQFT is its ability to
evaluate real-time, spatially or momentum resolved observables, such as charge
densities and currents, at finite times and in arbitrary background
fields, independently of a specific particle interpretation.

The approach introduced here provides the possibility of
real-time, operator-level simulation of vacuum
excitation in a genuinely curved spacetime using a numerically
tractable method. The approach therefore opens a complementary
pathway for investigating nonperturbative quantum field–theoretical phenomena
in curved spacetimes, with dynamical and operator-level control that
is difficult to obtain in existing numerical frameworks. In particular, we
study the Dirac field in a genuinely curved
$1\!+\!1$-dimensional spacetime and arrive at the well-known result that spacetime curvature
alone -- without electromagnetic fields -- can induce quench-driven vacuum
excitation and fermion--antifermion pair production with respect to a chosen
mode decomposition, visible in the real-time evolution of the vacuum state.

Our formalism integrates the covariant Dirac equation, the spin
connection, and curvature-dependent Hamiltonian into a numerically
stable operator-splitting scheme. This places our approach at the
intersection of semiclassical QFT in curved backgrounds, strong-field
QED numerics, and analogue-gravity simulations. The resulting framework
provides a complementary, nonperturbative tool to simulate fields in arbitrary time-dependent geometries, accounting for the dynamics inside localized curvature profiles. Although we follow the common approach of reducing the dimensionality of the problem to lower computational cost, the method developed is not restricted the 2-dimensional geometries studied here

Dilaton gravity
\cite{mandal,dilaton_grumiller_1,dilaton_frolov_1,dilaton_frolov_2}
provides a string-inspired framework \cite{mandal} in which such a
dimensional reduction can be carried out while retaining essential
gravitational features. In the present work, the spacetime metric is not
assumed to be a solution of a specific dilaton gravity model; rather, it
is inspired by dilaton-type geometries, and is chosen to
provide a smooth and nonsingular curved background suitable for
studying vacuum excitation and particle creation within the CQFT
framework.

The paper is organized as follows. Section~II reviews the CQFT formalism
in flat spacetime and the split-operator implementation.
In Sec.~III we extend this framework to a $1\!+\!1$-dimensional
curved background described by a smooth, localized Gaussian deformation
of flat spacetime, and derive a Hermitian Hamiltonian and its
unitary time-evolution operator. Section~IV presents numerical results
for the real-time evolution of the fermionic vacuum and the associated
pair-production observables. Finally, Sec.~V discusses the physical
interpretation and limitations of the particle-number definition used
here and outlines extensions to more realistic
measurement scenarios and additional external fields.

\section{Formalism of computational quantum field theory}

We begin with a brief review of the CQFT approach for electromagnetic interactions on a flat spacetime before then adapting this formalism to a curved manifold. As such we will present the Dirac equation in the context of both relativistic quantum mechanics and quantum field theory before outlining the connection between the two.

\subsection{First-quantized Dirac theory}

In flat spacetime, the Dirac equation for a relativistic particle in the presence of an external
electromagnetic four-potential $A_\mu(x)$ reads
\begin{align}
	\big(i\gamma^\mu D_{\mu}(x) - m \big)\psi(x) &= 0 ,\\
    D_{\mu} &= \partial_\mu - i A_\mu
\end{align}
where natural units $\hbar=c=e=1$ are used throughout this work.
Here $x^\mu=(t,\mathbf{x})$ denotes the spacetime coordinates with
$\mu=0,1,2,3$ and $A_\mu(x)$ represents a prescribed (fixed and classical) external electromagnetic
background. 
In a chosen inertial frame, this equation can be rewritten in Schrödinger form as
\begin{align}
\label{eq:Ham}
    i \partial_t \psi(t,\mathbf{x})
    &= \big(\hat{H}_{fr} 
    + \hat{H}_{em}\big)\, \psi(t,\mathbf{x}), \\
    \hat{H}_{em} &=  \gamma^0 \gamma^\mu A_\mu
\end{align}
where $\hat{H}_{fr}$ is the free Dirac Hamiltonian and $\hat{H}_{em}$ represents the electromagnetic interaction with the field.

We now specialize to fermionic field theory in $1+1$ spacetime
dimensions. In this case one may choose
the following representation of the Clifford algebra,
\[
\gamma^0=\sigma_3, \qquad \gamma^1=i\sigma_2 ,
\]
which leads to the free Hamiltonian
\begin{equation}
\hat{H}_{fr} = \hat{p}\, \sigma_1 + m \sigma_3 .
\end{equation}
From now on, $x$ will denote solely the spatial coordinate and $t$ the temporal coordinate in the laboratory frame. The first-quantized version of this theory -- or the one-particle theory -- has, of course, various well-known issues. However, we presented it here because the Hamiltonian in equation (\ref{eq:Ham}) will reappear when we reformulate the second-quantized theory below.

In the single particle theory, the Dirac equation is considered as applying to the wavefunction, $\psi(x)$. The Hilbert space is spanned by eigenstates of $\hat{H}_{fr}$, which are easily expressed in position space in terms of the familiar (2D) Dirac spinors:
\begin{equation*}
	\begin{split}
		\nu_{p}(x) 
		&= N(p)
		\begin{pmatrix}
			1 \\
			\dfrac{p}{m + E_p}
		\end{pmatrix}\, e^{-i p  x},\\
		\omega_{p}(x)
		&= N(p)
		\begin{pmatrix}
			-\dfrac{p}{m + E_p} \\
			1
		\end{pmatrix} \, e^{i p x},
	\end{split}
\end{equation*}
where $E_p=\sqrt{p^2+m^2}$ and $N(p)$ is a normalization factor.
In the above, $\nu_{p}(x)$ and $\omega_{p}(x)$ are the positive and negative frequency eigenfunctions of the one-particle Hamiltonian corresponding to momentum "p", $E_p = \sqrt{c^2 p^2 + m^2 c^4}$, $x$ is the spatial coordinate, and $N(p)$ is a normalization factor.

\subsection{Second quantization and QFT observables}

In the second-quantized theory, the fermionic field is promoted to an operator, expanded as
as
\begin{equation}
	\label{psit}
	\hat{\psi}(t,x)
	= \int dp \left(
	\hat{b}_p(t)\, v_p(x)
	+ \hat{d}^\dagger_p(t)\, w_p(x)
	\right),
\end{equation}
where $v_p(x)$ and $w_p(x)$ are the positive- and negative-energy solutions of the free Dirac equation, and
$\hat{b}_p(t)$ and $\hat{d}_p(t)$ are the fermionic annihilation operators
for particles and antiparticles, respectively -- these obey the standard anti-commutation relations for creation and annihilation operators and are used to build the Fock space of multi-particle states of the field.

The quantum-field-theoretical Hamiltonian is
\begin{equation}
	\hat{\mathcal{H}}
	= \hat{\psi}^\dagger(t,x)\, \hat{H}\, \hat{\psi}(t,x),
\end{equation}
where $\hat{H}$ is the corresponding first-quantized Hamiltonian in (\ref{eq:Ham}).
The connection between the second-quantized approach and the first-quantized Schrödinger form presented above is that it has been shown \cite{grobe_arb} that the Heisenberg equation of motion for the field operator,
\begin{equation}
	i\partial_t \hat{\psi}(t,x)
	= [\hat{\psi}(t,x),\hat{\mathcal{H}}],
\end{equation}
can be written (exactly) similarly to the single-particle evolution equation
\begin{equation}
	i\partial_t \hat{\psi}(t,x)
	= \hat{H}\, \hat{\psi}(t,x).
\end{equation}
Here, the Hamiltonian $\hat{H}$ acts on the spinor-valued mode functions
appearing in the expansion of the field operator. Consequently, the time evolution of the field operator can be generated by an operator, $\hat{U}(t)$, interpreted in the first-quantized setting, 
\begin{equation}
	\hat{\psi}(t,x)
	= \hat{U}(t)\, \hat{\psi}(0,x) =  e^{-i \hat{H} t}\, \hat{\psi}(0,x),
\end{equation}
where the final equality holds for stationary one-particle Hamiltonians. 

The time dependence of the creation and annihilation operators is
obtained by re-expanding the evolved field operator in a fixed reference
basis, chosen according to the physical context of the measurement. In
the presence of external electromagnetic fields that vanish
asymptotically in space or in time\footnote{The extension to gauge potentials that are asymptotically pure gauge is straightforward.}, a natural choice for this basis is the set of free Dirac eigenstates. Hence we can write

\begin{equation}
\label{creandanop}
	\begin{split}		
		\hat{b}_p(t)
		&= \int dp^\prime \left(
		U_{v_p v_{p^\prime}}(t)\, \hat{b}_{p^\prime}
		+ U_{v_p w_{p^\prime}}(t)\, \hat{d}^\dagger_{p^\prime}
		\right),\\
		\hat{d}_p^\dagger(t)
		&= \int dp^\prime \left(
		U_{w_p v_{p^\prime}}(t)\, \hat{b}_{p^\prime}
		+ U_{w_p w_{p^\prime}}(t)\, \hat{d}^\dagger_{p^\prime}
		\right),
	\end{split}
\end{equation}
where, for example, our matrix notation means 
\[
U_{w_p v_{p^\prime}}(t)
= \int dx\, v_p^\dagger(x)\, \hat{U}(t)\, w_{p^\prime}(x),
\]
and $\hat{b}_{p}$, $\hat{d}_{p}$ are the annihilation operators in the asymptotic past (in our scenario of a curvature quench this is equivalent to taking the operators at $t=0$), and similarly for creation operators. So the time-evolved operators define a Fock space in terms of the creation and annihilation operators in the asymptotic past.

With these identifications, the number-density operators associated with positive- and
negative-energy states are defined as
\begin{equation}
\label{rhoplusandminus}
	\begin{split}
		\hat{\rho}_{+}(t,x)
		&= \iint dp\, dp^{\prime}\,
		\hat{b}_p^\dagger(t)\, \hat{b}_{p^\prime}(t)\,
		v_{p}^\dagger(x)\, v_{p^\prime}(x),\\
		\hat{\rho}_{-}(t,x)
		&= \iint dp\, dp^{\prime}\,
		\hat{d}_p^\dagger(t)\, \hat{d}_{p^\prime}(t)\,
		w_{p^\prime}^\dagger(x)\, w_{p}(x).
	\end{split}
\end{equation}
Note that this definition implicitly contains the identification of particle states according to the time-evolved operators. The momentum spectrum of the created fermions (respectively anti-fermions) is determined by the number operators corresponding to each particle momentum,
\begin{equation}
\label{rhop}
	\begin{split}
		\hat{\rho}_{p+}(t,p)
		&= 
		\hat{b}_p^\dagger(t)\, \hat{b}_{p}(t)\,\\
        \hat{\rho}_{p-}(t,p)
		&= 
		\hat{d}_p^\dagger(t)\, \hat{d}_{p}(t).
	\end{split}
\end{equation}
Similarly, the charge-density operator is given by
\begin{equation}
	\hat{\rho}(t,x)
	= \hat{\psi}^\dagger(t,x)\, \hat{\psi}(t,x).
\end{equation}
Inserting the mode expansion only the particle and antiparticle (``diagonal'') terms
will survive once we take vacuum expectation values below. Using the fermionic anti-commutation relations, one finds 
\begin{equation}
	\begin{split}
		\hat{\rho}(t,x)
		&= \iint dp\, dp^\prime\,
		\hat{b}_p^\dagger(t)\, \hat{b}_{p^\prime}(t)\,
		v_p^\dagger(x)\, v_{p^\prime}(x)\\
		&\quad
		- \iint dp\, dp^\prime\,
		\hat{d}_{p^\prime}(t)\, \hat{d}^\dagger_p(t)\,
		w_p^\dagger(x)\,w_{p^\prime}(x) + \ldots \\
		&= \hat{\rho}_{+}(t,x) - \hat{\rho}_{-}(t,x) +\ldots.
	\end{split}
\end{equation}
where the omitted terms will drop out below.

To determine a number or charge density of created pairs in the Heisenberg picture, we determine the vacuum expectation values of $\hat{\rho}_+$,$\hat{\rho}_-$, and $\hat{\rho}_p$, obtaining, respectively:
\begin{equation}
\label{vev}
    \begin{split}
        \rho_+(t,x) &= \int dp \left| \int dp^\prime U_{v_p w_{p^\prime}}(t) v_p(x)\right| ^2   \\
        \rho_-(t,x)  &= \int dp \left| \int dp^\prime U_{w_p v_{p^\prime}} (t)w_p(x)\right| ^2  \\
        \rho_{p+}(t,p) &=  \int   dp^\prime \left|U_{w_p v_{p^\prime}} (t)\right| ^2  \\
        \rho_{p-}(t,p)&=   \int dp^\prime \left|U_{v_p w_{p^\prime}}(t)\right| ^2 \,. 
    \end{split}
\end{equation}
However, $\rho_{\pm}$ can only be interpreted unambiguously as corresponding to the \textit{physical} density of pairs created in the asymptotic limit $t \to \infty$, or at large spatial distances from variation in the electromagnetic field profile. This is because these expectation values depend explicitly on the choice of basis (through $\nu_{p}(x)$ and $\omega_{p}(x)$), or, physically, the choice of states that fill out the Fock space and define the multi-particle states.

\subsection{Split-operator approach to Computational Quantum Field Theory Evolution}
The CQFT  approach aims to evaluate observables at successive times during the real-time
evolution of the Dirac field operator generated by the first-quantized Hamiltonian with a finite spatial resolution. In order to do so, we discretize space to a one-dimensional lattice of width
$\Lambda$,
\begin{equation*}
	X = \left\{
	- \frac{N}{2}\delta x,\,
	\left(- \frac{N}{2} +1 \right)\delta x,\,
	\ldots,\,
	\frac{N}{2}\delta x
	\right\},
\end{equation*}
with lattice spacing $\delta x = \Lambda/(N+1)$.
The corresponding reciprocal-space lattice is
\begin{equation*}
	P = \left\{
	- \frac{N}{2}\delta p,\,
	\left(- \frac{N}{2} +1 \right)\delta p,\,
	\ldots,\,
	\frac{N}{2}\delta p
	\right\},
\end{equation*}
where $\delta p = 2\pi/\Lambda$.

The background scalar potential $V(x)$ and vector potential $A(x)$ are
evaluated on the position-space lattice, while the free Hamiltonian
$\hat{H}_{fr}$ is most conveniently evaluated in momentum space. For each momentum
$p=n\delta p$, with
$n\in\{-\tfrac{N}{2},-\tfrac{N}{2}+1,\ldots,\tfrac{N}{2}\}$, the (first-quantized) positive-
and negative-energy basis spinors are the momentum space eigenstates of $H_{fr}$:
\begin{equation*}
	\begin{split}
		\langle p|v_p\rangle
		&= N(p)
		\begin{pmatrix}
			1 \\
			\dfrac{p}{m + E_p}
		\end{pmatrix},\\
		\langle p|w_p\rangle
		&= N(p)
		\begin{pmatrix}
			-\dfrac{p}{m + E_p} \\
			1
		\end{pmatrix}.
	\end{split}
\end{equation*}

The time-evolution operator over a short time step $\delta t$ is written
in symmetrized (second-order) split-operator form as
\begin{equation}
    \hat{U}(\delta t)
    = \hat{U}_{em}\!\left(\frac{\delta t}{2}\right)
      \hat{U}_{fr}(\delta t)
      \hat{U}_{em}\!\left(\frac{\delta t}{2}\right),
\end{equation}
where $U_{fr}$ and $U_{em}$ denote the evolution operators associated with
the free Hamiltonian and the electromagnetic interaction, respectively.
The matrix element of the free evolution operator is given explicitly by
\begin{equation}
	\langle p' \vert \hat{U}_{fr}(\delta t) \vert p\rangle
	= \langle p' \vert  e^{- i \hat{H}_{fr} \delta t}
	\vert p\rangle= \begin{pmatrix}
		U_{11}& U_{12}\\
		U_{21}& U_{22}
	\end{pmatrix} \delta(p-p'),
\end{equation}
with matrix elements
\begin{equation}
	\begin{split}
		U_{11} &= \cos(E_p\delta t) - i \frac{m }{E_p} \sin(E_p \delta t),\\
		U_{12} &= -i \frac{ p}{E_p} \sin(E_p \delta t),\\
		U_{21} &= U_{12},\\
		U_{22} &= \cos(E_p\delta t) + i \frac{m }{E_p} \sin(E_p \delta t).
	\end{split}
\end{equation}
The interaction step is applied in position space and reads
\begin{equation}
 \langle x' \vert  \hat{U}_{em}(\tau)\vert x\rangle
   = \exp\!\left[- i \left(V(x) + \sigma_1 A(x)\right)\tau\right] \delta(x - x').
\end{equation}
Transition amplitudes such as
$U_{w_p v_p}\equiv \langle w_p|\hat{U}|v_p\rangle$,
$U_{w_p,w_p'}\equiv \langle w_p|\hat{U}|w_p'\rangle$ and so on, are evaluated numerically. With these in hand, we can then evaluate the created pair number and / or charge densities by evaluating (\ref{vev}) on the discretised momentum space lattice.

The same CQFT approach presented in this section can be used to calculate the numbers and number
densities of created boson--antiboson pairs and to study the propagation
of bosonic wave packets in strong electromagnetic fields that are
supercritical with respect to the bosonic mass \cite{oursmicrocausality,wagner2010bosonic}. This approach has also been used to study the dynamics of Klein tunneling
\cite{ours_qft_klein} and regular tunneling of both fermionic and bosonic
fields \cite{oursfullycausal}, both where the background field vanishes
asymptotically and inside the potential barrier through which the
wavepacket tunnels. For this reason,
the same projection onto the asymptotic mode decomposition as in Eq.~\eqref{creandanop} was used in those
works.

However, in other contexts where the background field is
time dependent, different re-expansions can be employed. For example,
one may use a basis that diagonalizes the Hamiltonian instantaneously. This yields the same particle numbers as the adiabatic particle number. As discussed in the introduction, the adiabatic particle number can exhibit intermediate values that are orders of magnitude greater than the asymptotic one converged to at late times. Using different bases for the mode expansion, say the basis that diagonalizes the Hamiltonian at $ t \rightarrow \infty $, would result in different numbers at finite times. The dependence of particle number, calculated at intermediate times, on the quantization scheme is also observed in phase-space approaches to pair production (see, e.g., \cite{ildertondhw}).

As an illustration, consider Sauter's time-dependent field,
\begin{equation}
\label{sauter_step}
	a(t) = \frac{E_0}{\omega}\left(1+\tanh(\omega t)\right)\,.
\end{equation}
One option is to use the basis that instantaneously diagonalizes the Hamiltonian given in
Eq.~\eqref{eq:Ham}. Alternatively, one may project on to the Hamiltonian at asymptotically late times, for which $a(t) \to a^{\infty}$ becomes constant (pure gauge) and recover
the adiabatic particle
number \cite{ildertonadia}. In the late time limit, the pair creation rates agree. 

\begin{figure}[tbp]
\centering
\includegraphics[width=0.4\textwidth]{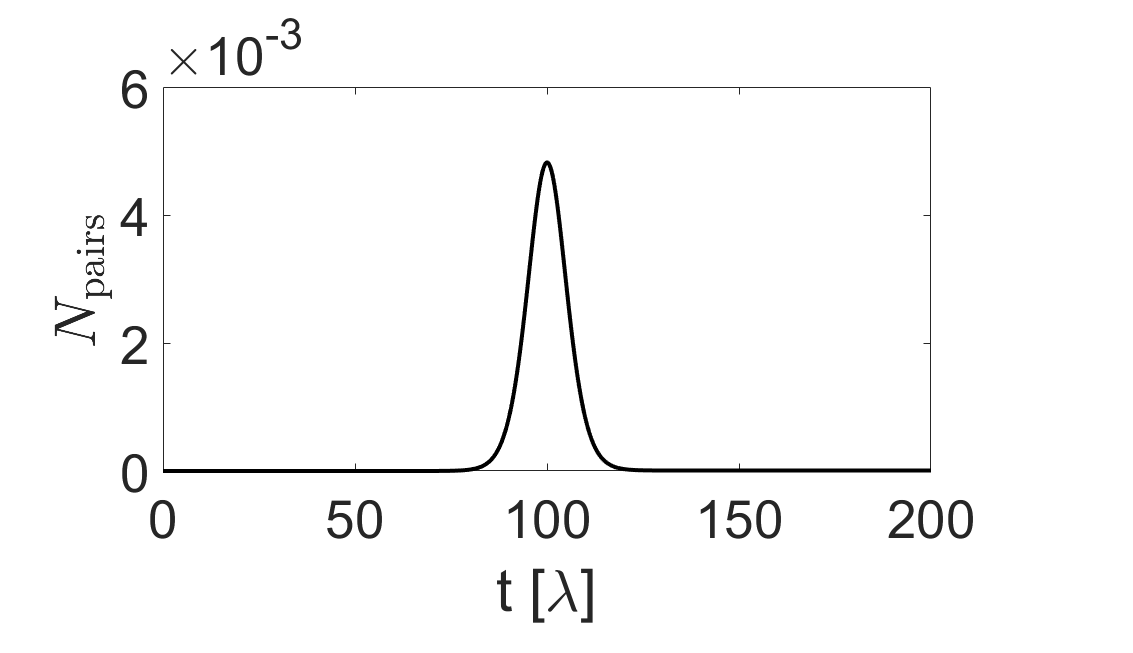}
\includegraphics[width=0.4\textwidth]{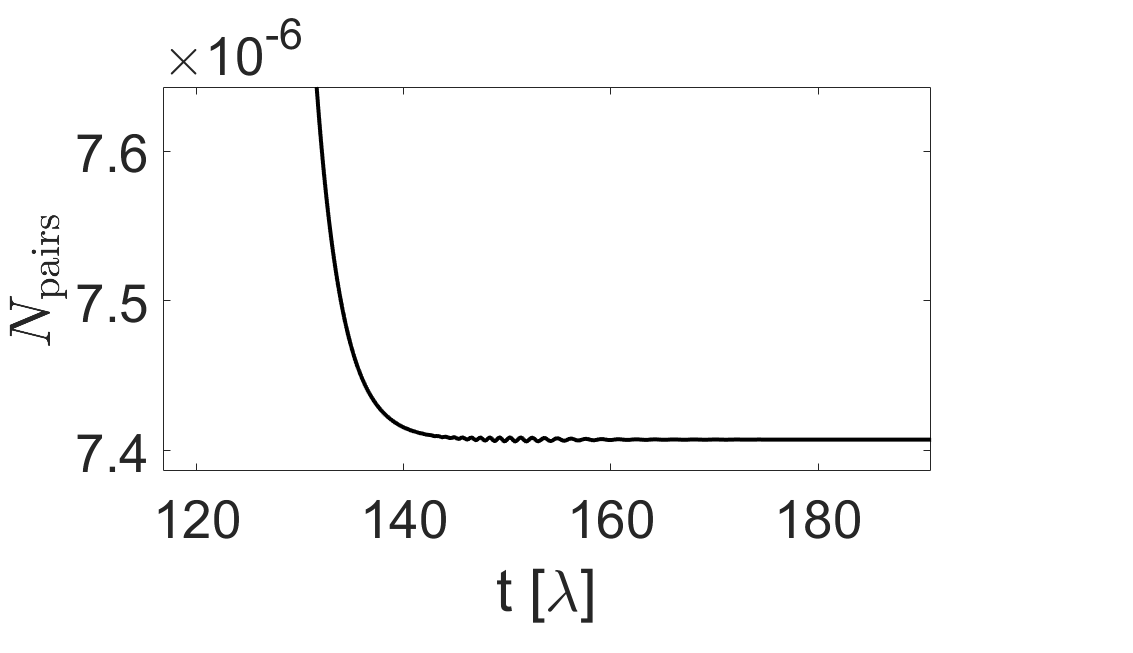}
\caption{Number of created fermion--antifermion pairs for a Sauter step given by Eq.~\eqref{sauter_step}, with $E_0=1/4$, and $\omega=0.1$. The upper panel shows the time evolution over a duration of $200 \lambda$, while the lower panel is a zoom of the late-time region. The obtained asymptotic number is $N \simeq 7.4 \times 10^{-6}$. The calculation is performed using CQFT on a lattice of width $100\,\lambda$ with $2^{11}$ sites and time step $\delta t = 0.001 \lambda$ increasing the number of lattice sites or reducing the time step does not change the curves at the resolution shown. The asymptotic number of created pairs agrees with the asymptotic adiabatic particle number~\cite{ildertonadia}.}
\label{adiafig}
\end{figure}

In Fig. (\ref{adiafig}) we show the number of created pairs as determined by CQFT using the basis that instantaneously diagonalizes the Hamilotinian given by Eq. \eqref{eq:Ham} with the background corresponding to the Sauter's field given by Eq. \eqref{sauter_step}, meaning that in the asymptotic limit we are defining pair creation by projecting (instead of Eq.~\eqref{creandanop}) onto the basis of eigenstates of the Hamiltonian which is related to the free basis by Eq.

\begin{equation}
\begin{split}
v^{(a_\infty)}_{p}(t,x) &= e^{-i a_\infty x}\, v_p(x),\\
w^{(a_\infty)}_{p}(t,x) &= e^{-i a_\infty x}\, w_p(x),
\end{split}
\end{equation}
or equivalently by
\begin{equation}
\begin{split}
v^{(a_\infty)}_{p}(t,x) &\equiv v_{p+a_\infty}(x),\\
w^{(a_\infty)}_{p}(t,x) &\equiv w_{p+a_\infty}(x).
\end{split}
\end{equation}
This, in fact, means that we end up projecting onto the same (pure gauge) basis used in \cite{ildertonadia}.

\section{Computational quantum field theory in curved spacetimes}
To adapt CQFT to curved spacetime we will start with the Dirac equation and show that (at least in (1+1)-dimensional spacetime) it can be recast into Schrödinger form. We will then show that the first-quantized quantum theory can be used to generate time evolution for the second-quantized theory as for electromagnetic interactions, although achieving this will require a field redefinition. For details on our curved space conventions, see Appendix \ref{App:Curved}.

In curved spacetime, the free massive Dirac equation takes the covariant form (see (\ref{diraceqapp}))
\begin{equation}
	i\gamma^\nu(x) \nabla_\nu \psi(x) - m \psi(x) = 0 ,
\end{equation}
where $\nabla_\nu$ denotes the spinor covariant derivative. Using the
vielbein formalism, whereby we set $g_{\mu\nu}(x) = e_{\mu}{}^{a}(x) e_{\nu}{}^{b}(x) \eta_{ab}$ and $\gamma^{\mu} = e^{\mu}{}_{a}\gamma^{a}$, this equation can be written explicitly as
\begin{equation}
	\label{diracgeneral}
	i \tensor{e}{^\nu_a}(x)\, \gamma^a
	\left( \partial_\nu + \Omega_\nu(x) \right)\psi(x)
	- m \psi(x) = 0 ,
\end{equation}
where $\tensor{e}{^\nu_a}$ are the vielbeins and $\Omega_\nu$ is the spinor
connection (Fock--Ivanenko coefficient) detailed further below.

Throughout this section, we distinguish between curved spacetime
coordinates $(\tau,\xi)$ and local Lorentz (tangent-space) coordinates
$(t,x)$, and we will use the Greek letters corresponding to the curved coordinates instead of their respective numerical indices when citing the connections and tetrads. For example, we will use $\Gamma^\xi_{\tau \xi}$ and $e^{\tau}{}_{0}$ rather than $\Gamma^{1}_{0 1}$ and $e^{0}{}_{0}$ so that the curved coordinate variable is indicated explicitly.

\subsection{Background geometry}
We consider a static $1\!+\!1$-dimensional spacetime that is flat Minkowski space for $\tau < 0$. We introduce a deformation for $\tau \geqslant 0$, whereby the manifold becomes curved, with line element
\begin{equation}
	\label{metricschwa}
	ds^2
	= \alpha(\xi)\, d\tau^2
	- \frac{d\xi^2}{\alpha(\xi)}\,  ,
\end{equation}
where
\begin{equation}
\label{defo}
\alpha(\xi)
= 1 - \beta e^{-\xi^2/r_0^2} \,.
\end{equation}
Here, then, $r_0$ controls the width of the Gaussian deformation of flat
spacetime and $\beta$ control its strength. In order to avoid coordinate singularities and keep the metric signature fixed, we choose to take $\beta$ to be smaller than unity.

The metric \eqref{metricschwa} is chosen as a smooth, asymptotically flat
deformation of flat spacetime that is free of curvature singularities.
Such regular background geometries are particularly well suited for
numerical investigations of quantum field dynamics in curved spacetime.
Although no assumption is made regarding the microscopic gravitational origin of
this geometry, the metric Eq.~\eqref{metricschwa} is inspired by a classical solution of a 2-dimensional string theory \cite{mandal,dilaton_frolov_1} where the metric takes the same form as in Eq.~\eqref{metricschwa} with $\alpha(\xi) = 1- \beta e^{-Q \xi}$.

This restriction to smooth spacetimes is motivated by fundamental
considerations in quantum field theory on curved backgrounds. In
spacetimes containing curvature singularities or lacking global
hyperbolicity, the quantum Hamiltonian governing field evolution
generally fails to be essentially self-adjoint. As a consequence, quantum
time evolution is not uniquely defined unless additional boundary
conditions are specified at the singularity, leading to ambiguities in
the dynamics \cite{wald1980}. By working with a regular background
geometry that is asymptotically flat, these issues are avoided and a well-defined, unitary time
evolution will ensue (see below).

The non-vanishing Christoffel symbols associated with the metric
\eqref{metricschwa} are (see Appendix \ref{Gammas})
\begin{equation}
	\begin{split}
		\Gamma^{\tau}_{\xi \tau}
		&= \Gamma^{\tau}_{\tau \xi}
		= \frac{\alpha'(\xi)}{2\alpha(\xi)}, \\
		\Gamma^{\xi}_{\tau\tau}
		&= \frac{1}{2}\alpha(\xi)\alpha'(\xi), \\
		\Gamma^{\xi}_{\xi\xi}
		&= -\frac{\alpha'(\xi)}{2\alpha(\xi)} ,
	\end{split}
\end{equation}
where a prime denotes differentiation with respect to $\xi$. From here one finds a non-zero Riemann curvature tensor characterised by the Ricci scalar,
\begin{equation}
\label{ricci}
    R(\xi) = \alpha''(\xi)\ = \frac{2\,\beta}{r_0^{4}}
\, e^{-\xi^{2}/r_0^{2}}
\left( r_0^{2} - 2\,\xi^{2} \right), \qquad \tau > 0
\end{equation}
(with the Kretschmann scalar being the square of this). This confirms that the metric remains flat at asymptotic spatial distance (recall that in (1+1)-dimensional spacetime the Ricci tensor is given by $R_{\mu\nu}(\xi) = \frac{1}{2}R(\xi)g_{\mu\nu}(\xi)$).

Before proceeding let us address the question of continuity of the metric at $\tau = 0$. Of course a smooth switch-on could be implemented by introducing a time-dependent coefficient, e.g. $\alpha(\tau,\xi)=1 - s(\tau)\,e^{-\xi^2/r_0^2},$ where $s(\tau)$ interpolates from $0$ to $\beta$. This would introduce additional time-derivative terms in the Dirac equation arising from the time dependence of the tetrads and spin connection. For simplicity, here we consider a sudden quench  allowing the lattice spacing to regulate the discontinuity at $\tau = 0$. Terms proportional to $\delta(\tau)$, which can arise from derivatives of the step function $\partial_\tau \Theta(\tau)$ in a global metric description, are instead encoded in matching conditions at $\tau = 0$, which in the present formulation reduce to continuity of the rescaled field,
\begin{equation}
\chi(0^+,\xi) = \chi(0^-,\xi).
\end{equation}
The post-quench evolution for $\tau > 0$ is then generated by Eq.~(\ref{hamcur}).

\subsection{Vielbeins and spin connection}

We choose the diagonal vielbein (see Appendix \ref{tetrads})
\begin{equation}
	\label{vielbein}
	\tensor{e}{^\tau_0} = \frac{1}{\sqrt{\alpha(\xi)}},
	\qquad
	\tensor{e}{^\xi_1} = \sqrt{\alpha(\xi)},
\end{equation}
with all other components vanishing.

The only non-vanishing components of the spin connection are (see Appendix \ref{omegas} and \ref{omegaantisy})
\begin{equation}
	\omega^{0}_{1\tau}
	= \omega^{1}_{0\tau}
	= \frac{\alpha'(\xi)}{2},
\end{equation}
while $\omega^a_{b\xi}=0$. Then the spinor connection for the matter field is given by (see Appendix \ref{Omegas})
\begin{equation}
	\Omega_\nu
	= -\frac{i}{4}\omega_{ab\nu}\sigma^{ab},
	\qquad
	\sigma^{ab}
	= \frac{i}{2}[\gamma^a,\gamma^b].
\end{equation}
Using the $1\!+\!1$-dimensional representation
$\gamma^0=\sigma_3$ and $\gamma^1=i\sigma_2$, one finds
\begin{equation}
	\label{Omega}
	\Omega_\tau
	= \frac{\alpha'(\xi)}{4}\, \sigma_1,
	\qquad
	\Omega_\xi = 0 .
\end{equation}

\subsection{Dirac equation and Hamiltonian formulation}
Inserting Eqs.~\eqref{vielbein} and \eqref{Omega} into
\eqref{diracgeneral}, the Dirac equation in our curved
spacetime becomes (for $\tau > 0$)
\begin{equation}
	\label{dirac2d}
    \begin{split}
	\frac{i}{\sqrt{\alpha(\xi)}}\, \sigma_3
	\left(
	\partial_\tau
	+ \frac{\alpha'(\xi)}{4}\, \sigma_1
	\right)\psi(\tau,\xi)
	\\ -
	\left(
	\sqrt{\alpha(\xi)}\, \sigma_2\, \partial_\xi
	+ m 
	\right)\psi(\tau,\xi)
	= 0 .
    \end{split}
\end{equation}
The covariant Dirac current $j^\mu=\bar\psi\gamma^\mu\psi$ satisfies
$\nabla_\mu j^\mu=0$ and defines a conserved inner product on a Cauchy surface of constant $\tau$). Taking the curvature into account, this inner product is defined by 
\begin{equation}
	(\psi_1,\psi_2)
	= \int d\xi \frac{1}{\sqrt{\alpha(\xi)}}\,\psi_1^\dagger\psi_2 .
\end{equation}
Our aim is to rewrite the equation of motion for the field in Schrödinger form so as to develop a first-quantized representation of its time evolution. Nevertheless, the
Dirac Hamiltonian obtained from Eq.~\eqref{dirac2d} is not manifestly Hermitian
with respect to the flat $L^2(d\xi)$ norm due to the $\xi$-dependence of the
vielbeins. We therefore perform the field redefinition
\begin{equation}
\label{rescaling}
    \chi(\tau,\xi) =  \alpha^{-\frac{1}{4}}(\xi) \psi (\tau,\xi)    
\end{equation}
 which yields a Hamiltonian that is Hermitian with
respect to the standard flat inner product
\begin{equation}
(\chi_1,\chi_2)=\int d\xi\,\chi_1^\dagger\chi_2.
\end{equation}

In the first-quantized setting, the Dirac equation, Eq~\eqref{dirac2d}, can then be rewritten in the Schrödinger form in terms of the rescaled field,
\begin{equation}
\label{dirac_scaled}
	i \partial_\tau \chi(\tau,\xi)
	= \hat{H}\, \chi(\tau,\xi),
\end{equation}
where $\tau$ denotes the curved-spacetime time coordinate associated with
the background metric \eqref{metricschwa}. The Hamiltonian is (for $\tau > 0$)
\begin{equation}
	\label{hamcur}
	\hat{H}
	= \frac{1}{2}\{\alpha(\xi),\hat{p}\}\, \sigma_1
	+ \sqrt{\alpha(\xi)}\, m  \sigma_3.
\end{equation}
The anticommutator ensures Hermiticity of the Hamiltonian operator.

It is convenient to decompose $\hat{H}$ into the free flat-spacetime
Hamiltonian $\hat{H}_{fr}$ and curvature-induced corrections:
\begin{align}
	\label{hamcurcom1}
	\hat{H}
	&= \hat{H}_{fr} + \hat{H}_{gr}\\
	\hat{H}_{gr} &= \frac{1}{2}
	\{\alpha(\xi)-1,\hat{p}\}\, \sigma_1
	+ \bigl(\sqrt{\alpha(\xi)}-1\bigr)m \sigma_3.
\end{align}
For the metric \eqref{metricschwa}, this yields
\begin{equation}
	\label{hamcurcom2}
	\hat{H}
	= \hat{H}_{fr}-
	\frac{1}{2}
	\{\beta e^{-\xi^2/r_0^2},\hat{p}\}\, \sigma_1
	+ \bigl(\sqrt{1-\beta e^{-\xi^2/r_0^2}}-1\bigr)m \sigma_3.
\end{equation}
 the free part of the Hamiltonian can be used to define the  states of the theory corresponding to flat space.

It is important to note, however, that a rescaling analogous to that introduced in
Eq.~\eqref{rescaling} can be carried out in the general $1\!+\!1$-dimensional
case once the spacetime metric is written in conformally flat form.
Indeed, every two-dimensional Lorentzian metric is locally conformally
flat. Consequently, a metric of the general form
\begin{equation}
ds^2 = \alpha_1(\tau,\xi)\, d\tau^2 - \alpha_2(\tau,\xi)\, d\xi^2
\end{equation}
can always be brought, by a suitable local coordinate transformation, to
the form  (See App.~\ref{app:conformal_flatness})
\begin{equation}
ds^2 = \alpha(\tau',\xi')\left(d\tau'^2 - d\xi'^2\right).
\end{equation}
In these conformal coordinates, the spin connection components can be
computed straightforwardly. In the representation
$\gamma^0=\sigma_3$ and $\gamma^1=i\sigma_2$, one finds
\begin{equation}
\Omega_{\tau'}
= \frac{\partial_{\xi'}\alpha(\tau',\xi')}{4\,\alpha(\tau',\xi')}\,\sigma_1,
\qquad
\Omega_{\xi'}
= \frac{\partial_{\tau'}\alpha(\tau',\xi')}{4\,\alpha(\tau',\xi')}\,\sigma_1.
\end{equation}
Introducing the rescaled spinor field
\begin{equation}
\label{rescaling_confo}
\phi(\tau',\xi')=\alpha^{1/4}(\tau',\xi')\,\psi(\tau',\xi'),
\end{equation}
the Dirac equation assumes the Schrödinger-like form
\begin{equation}
i\,\partial_{\tau'}\phi(\tau',\xi')
=
- i\,\sigma_1\,\partial_{\xi'}\phi(\tau',\xi')
+ m\,\sqrt{\alpha(\tau',\xi')}\,\sigma_3\,\phi(\tau',\xi').
\end{equation}
The corresponding Hamiltonian operator is manifestly Hermitian with
respect to the flat inner product, and the resulting time evolution is
unitary.

\subsection{Field quantisation and split-operator time evolution}
Now we quantise the fields, by setting
\begin{equation}
    \hat{\chi}(\tau, \xi) = \int dp \left( \hat{b}_p(\tau)\, v_p(\xi) 	+  \hat{d}^\dagger_p(\tau)\, w_p(\xi) \right),
\end{equation}
where $v_p(x)$ and $w_p(x)$ are the solutions of the free Dirac equations in the Minkowski flat spacetime.

As in flat space, the second-quantized Hamiltonian density is
\begin{equation}
    \hat{\cal{H}} = \hat{\chi}^\dagger(\tau,\xi)\hat{H}\hat{\chi}(\tau,\xi),
\end{equation}
and the Heisenberg equation for the time evolution of the field can, again, be written using the first-quantized Hamiltonian as
\begin{equation}
    i\partial_{\tau} \hat{\chi}(\tau, \chi) = \hat{H}  \hat{\chi}(\tau, \chi)\,.
\end{equation}
As such, we may employ the time evolution operator corresponding to the single-particle Hamiltonian.

The free evolution operator for $\hat{\chi}$ is identical to that obtained in the
flat-spacetime case discussed in Sec.~II, with the Minkowski time
coordinate $t$ replaced by the curved-spacetime time coordinate $\tau$:
\begin{equation}
	\hat{U}_{fr}(\delta\tau)
	= e^{-i\hat{H}_{fr}\delta\tau}.
\end{equation}
Its explicit matrix representation is therefore not repeated here.

Using a symmetric Baker--Campbell-Hausdorff split-operator scheme, the
full time-evolution operator is approximated to $\mathcal{O}(\delta \tau^{3})$ as
\begin{equation}
\label{Ucst}
	\hat{U}(\delta \tau)
	=
\hat{U}_{cur2}\!\left(\tfrac{\delta\tau}{2}\right)
	\hat{U}_{cur1}\!\left(\tfrac{\delta\tau}{2}\right)
	\hat{U}_{fr}(\delta\tau)
	\hat{U}_{cur1}\!\left(\tfrac{\delta\tau}{2}\right)
	\hat{U}_{cur2}\!\left(\tfrac{\delta\tau}{2}\right).
\end{equation}
Here $\hat{U}_{cur1}$ corresponds to the momentum-dependent curvature term
involving $\{e^{-\xi^2/r_0^2},\hat{p}\}$ and is evaluated most conveniently in momentum space,
\begin{equation}
	\hat{U}_{cur1}({\delta \tau})
	= \exp\!\left[
	- \frac{i}{2} \{\hat{f},\hat{p}\}
	\sigma_1\, {\delta\tau}
	\right],
\end{equation}
where 	$f(\xi)=\beta e^{-\xi^2/r_0^2}$ and
the momentum--space matrix elements of $\hat f$ are
\begin{equation}
\begin{split}
\langle p | \hat f | p' \rangle
&=
\int d\xi\,
\langle p | \xi \rangle
f(\xi)
\langle \xi | p' \rangle \\
&=
\frac{1}{2\pi}
\int d\xi\,
f(\xi)\,e^{i(p'-p)\xi}.
\end{split}
\end{equation}
On the other hand, $\hat{U}_{cur2}$ is applied in position space,
\begin{equation}
	\langle \xi' | \hat{U}_{cur2}({\delta \tau}) | \xi\rangle
	= \exp\!\left[
	- i g(\xi)\, {\delta\tau}
	\right] \delta(\xi - \xi'),
\end{equation}
with
\begin{equation}
	g(\xi)
	=
	\bigl(\sqrt{1-\beta e^{-\xi^2/r_0^2}}-1\bigr)m \sigma_3.
\end{equation}

The resulting time-evolution operator $\hat{U}$ is used in direct
analogy with the flat-spacetime case discussed in Sec.~II. In
particular, it is applied to evolve the quantized field in time and to
compute transition amplitudes between time-evolved single-particle states. These amplitudes are then used to construct the time-dependent creation and annihilation operators and to evaluate expectation values of quantum
field observables.

Since the system is initially prepared in the Minkowski vacuum prior to
the introduction of the spacetime curvature quench, the initial field
operator is the free Dirac field $\hat{\psi}(0,\xi)$ given by
Eq.~\eqref{psit} -- this provides the boundary condition for $\hat{\chi}(0, \xi)$. The subsequent time evolution can be described
equivalently either by the original Dirac equation
Eq.~\eqref{dirac2d} or by the rescaled equation
Eq.~\eqref{dirac_scaled}. We evolve the latter, because it yields a Hermitian Hamiltonian and
unitary time evolution in our first-quantized representation.

As in the flat-spacetime formulation, we project the time evolved field onto a reference basis. In the present work, this basis is chosen to be
the set of free Dirac eigenstates. This choice is natural for the
background geometry considered here, which is asymptotically flat, and
allows for a direct interpretation of particles and antiparticles in
terms of free asymptotic states. 

The \textbf{physical} densities are those corresponding to the conserved currents and can be expressed equivalently in terms of the field $\hat{\psi}(t,x)$ or $\hat{\chi}(t,x)$
\begin{equation}
\label{chdenopcur}
	\hat{\rho}_{ch}= \frac{1}{\sqrt{\alpha (\xi)}} \hat{\psi}^{\dagger} (\tau,\xi)\hat{\psi}(\tau,\xi) = \hat{\chi}^{\dagger} (\tau,\xi)\hat{\chi}(\tau,\xi)
\end{equation}

The number density operators for one particle positive energy and negative energy states $\rho_{+}$ are respectively:
\begin{equation}
	\begin{split}
		\hat{\rho}_{+}(\tau,\xi)
		&= \frac{1}{\sqrt{\alpha(\xi)}} \iint dp\, dp^{\prime}\,
		\hat{b}_p^\dagger(t)\, \hat{b}_{p^\prime}(t)\,
		v_{p}^\dagger(\xi)\, v_{p^\prime}(\xi),\\
        \hat{\rho}_{-}(\tau,\xi)
		&= \frac{1}{\sqrt{\alpha(\xi)}} \iint dp\, dp^{\prime}\,
		\hat{d}_p^\dagger(t)\, \hat{d}_{p^\prime}(t)\,
		w_{p}^\dagger(\xi)\, w_{p^\prime}(\xi),\\
	\end{split}
\label{nbdencur1}
\end{equation}
if these operators are derived from the field decomposition of $\psi$ and
\begin{equation}
	\begin{split}
		\hat{\rho}_{+}(\tau,\xi)
		&=  \iint dp\, dp^{\prime}\,
		\hat{b}_p^\dagger(t)\, \hat{b}_{p^\prime}(\tau)\,
		v_{p}^\dagger(\xi)\, v_{p^\prime}(\xi), \\
        	\hat{\rho}_{-}(\tau,\xi)
		&=  \iint dp\, dp^{\prime}\,
		\hat{b}_p^\dagger(t)\, \hat{b}_{p^\prime}(\tau)\,
		w_{p}^\dagger(\xi)\, w_{p^\prime}(\xi),
	\end{split}
\label{nbdencur2}
\end{equation}
if they are derived from the field decomposition of $\chi$.
The number density operators in the momentum space are given by
\begin{equation}
\label{rhop}
	\begin{split}
		\hat{\rho}_{p+}(\tau,p)
		&= 
		\hat{b}_p^\dagger(\tau)\, \hat{b}_{p}(\tau)\,\\
        \hat{\rho}_{p-}(\tau,p)
		&= 
		\hat{d}_p^\dagger(\tau)\, \hat{d}_{p}(\tau).
	\end{split}
\end{equation}
The number density of one-particle states defined with respect to the
Minkowski free basis as introduced in Eqs.~\eqref{nbdencur1} and \eqref{nbdencur2}
are basis dependent and correspond to excitations measured relative to the Minkowski vacuum (this can be shown by calculating the time evolution after turning off the spacetime curvature at any instant such the evolution continues with the free Minkowski space Hamiltonian only whereafter the number density remains constant). In contrast, the charge density operator given by Eq.~\eqref{chdenopcur} is a basis-independent observable. 

Using the time evolution operator Eq.~\eqref{Ucst}, one can compute the amplitudes $U_{w_p v_p}$, $U_{v_p v_p'}$ and so on. The number density of the created fermion--anti-fermion pairs $\rho_{\text{pairs}}$ and their spectrum $\rho_p(p)$, all calculated with respect to the flat spacetime vacuum, are then obtained by computing the vacuum expectation value of $\hat{\rho}_+(\tau,\xi)$ or $\hat{\rho}_-(\tau,\xi)$ given by Eqs.~\eqref{nbdencur1}. The vacuum expectation values are given by 

\begin{equation}
\label{vev_cur}
    \begin{split}
        \rho_+(\tau,\xi) &= \int dp \left| \int dp^\prime U_{v_p w_{p^\prime}}(\tau) v_p(\xi)\right| ^2   \\
        \rho_-(\tau,\xi)  &= \int dp \left| \int dp^\prime U_{w_p v_{p^\prime}} (\tau)w_p(\xi)\right| ^2  \\
        \rho_{p+}(\tau,p) &=  \rho_{p-}(\tau,p) =\int   dp^\prime \left|U_{w_p v_{p^\prime}} (\tau)\right| ^2 \\
        &=   \int dp^\prime \left|U_{v_p w_{p^\prime}}(\tau)\right| ^2 \,. 
    \end{split}
\end{equation}
while the expectation value of the charge density is vanishing everywhere, as no charge separation is taking place since the spacetime curvature does not couple to charge, and no electromagnetic field is included in the scenarios discussed in this work.
The number of pairs $N_{\text{pairs}}$ is then obtained by integrating $\rho_{\text{pairs}}$ over $\xi$ or by integrating $\rho_p(p)$ over p so that one obtains:
\begin{equation}
    N_{\text{pairs}} = \int dp \int dp^\prime \left| U_{v_p w_p'}(\tau) \right|^2
\end{equation}

It is important to note that in the framework of the CQFT presented in this work the background geometry can be time-dependent as one can introduce space- and time-dependent perturbations to the flat Minkowski vacuum. However, neither the backreaction nor the self-interaction of the matter field is accounted for.  The fermionic field is therefore evolved on the background geometry without accounting for gravitational backreaction or fermion self-interactions. The present formulation thus corresponds to quantum
field theory on a prescribed classical spacetime. Extensions of the method that incorporate backreaction and self-reaction effects and loop corrections will be considered in future work.


\section{Results and illustrations}
Starting from the field characterizing the  fermionic vacuum, we evolve it using the time evolution operator calculated in the curved spacetime characterized by the Gaussian deformation of the flat spacetime, Eq (\ref{metricschwa}) . We calculate the number densities and the spectrum of the created fermions in position space.

\begin{figure}[t]
\centering
\includegraphics[width=0.4\textwidth]{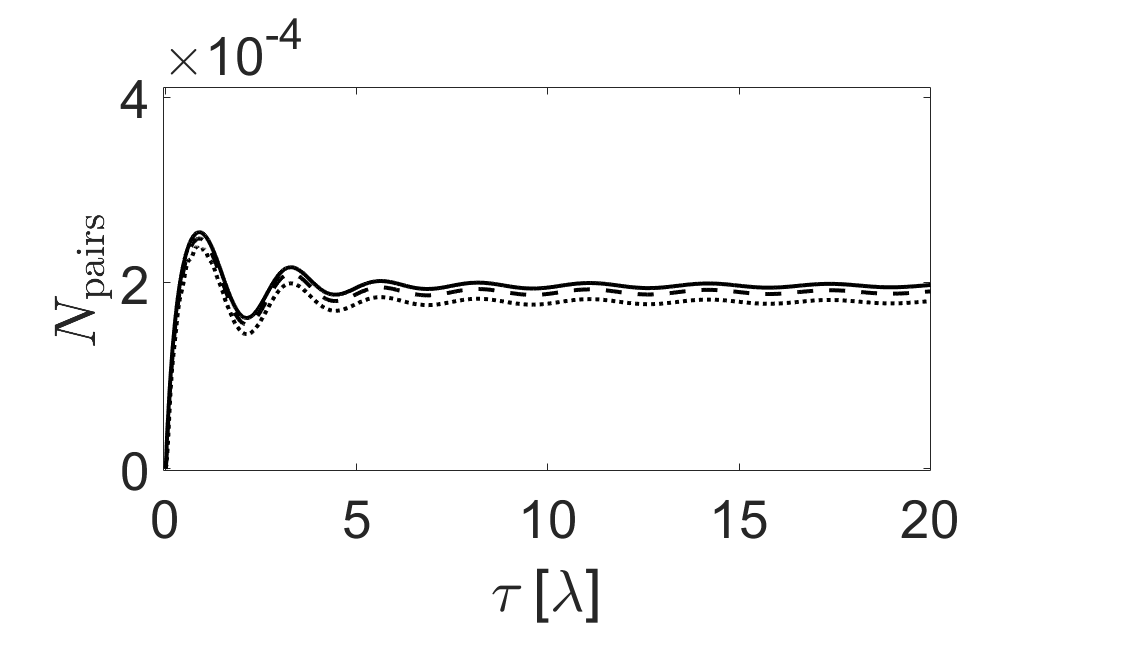}
\caption{Number of created pairs in the case of a Gaussian deformation of width $r=1 \lambda$ and lattice width $\Lambda = 100 \lambda$ for different lattice spacings corresponding to $N=2^9$ sites (dotted line),  $N=2^{10}$ (dashed line), $N=2^{11}$ (solid line). For greater values of $N$, the curves become indistinguishable in the plotting resolution. }
\label{nbs_conv}
\end{figure}

To assess numerical stability, we performed convergence tests with
respect to both the spatial lattice spacing $\Delta \xi$ and the time step
$\Delta \tau$. For sufficiently small $\Delta \xi$ and $\Delta \tau$ the results
for the particle number and spatial number density become indistinguishable
within the plotting resolution, indicating that the calculation is in the 
converged regime. Fig.~\ref{nbs_conv} shows the total number of 
fermion--antifermion pairs created for several lattice resolutions; convergence 
is reached for $\Delta \xi \leq 0.097 \lambda$ and $\Delta\tau \leq 0.01 \lambda$.

\begin{figure}[b]
\centering
\includegraphics[width=0.4\textwidth]{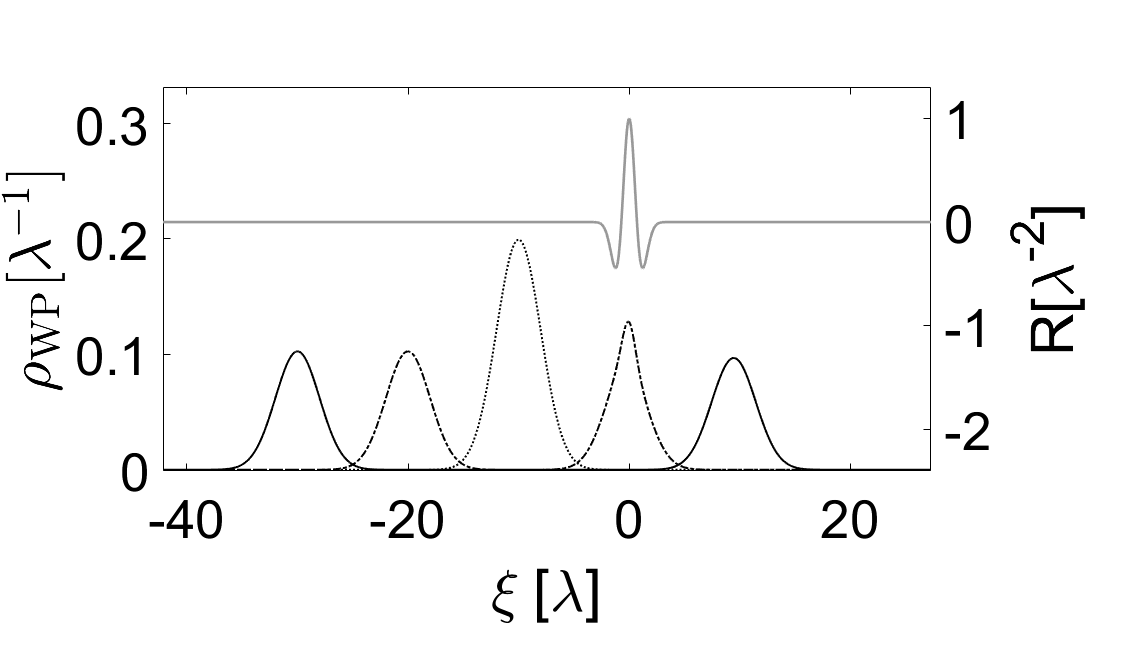}
\caption{The probability density of a a one-particle fermionic wave packet  evolved with the time evolution operator Eq.~\eqref{Ucst}. The initial wave packet is given by Eq.~\eqref{wpeq} and its corresponding density is shown by the dotted line. The dash-dotted line shows the density corresponding to the evolved wave packet at $t=10 \lambda$, and the solid line shows it at $t=15 \lambda$. The discretization uses a lattice of width $100 \lambda$ and of $10^{11}$ sites. The time step is $\delta_t = 0.01 \lambda$  }
\label{wp}
\end{figure}

As a complementary diagnostic we also monitored the norm of a Gaussian wave packet to test the unitarity of the evolution. Using the time evolution operator, Eq.~\eqref{Ucst}, we evolved a one-particle wave packet propagating through the curved spacetime. The initial wave packet was chosen to be

\begin{equation}
\label{wpeq}
\chi(0,x) =
\left( \frac{1}{2\pi\sigma^2} \right)^{1/4}
e^{-\frac{(x-x_0)^2}{4\sigma^2}}
e^{- i p_0 x}
\begin{pmatrix}
1 \\
0
\end{pmatrix},
\end{equation}
with $x_0 = -10 \lambda$, $p_0 = 1000 \lambda^{-1}$, and $\sigma=2 \lambda$. We found that the symmetrized operator splitting used in Eq.~\eqref{Ucst} preserves norm to sufficiently high accuracy; over the full simulation time of $20 \lambda$ the deviation 
$\epsilon(\tau)=\Vert\psi(t)\Vert^{2}-1$ remains below $3.4\times 10^{-13}$ for $\delta \xi = 0.0488 \lambda$ and $\delta \tau = 0.01 \lambda$ and it becomes smaller for smaller $\delta \xi$ and $\delta \tau$.
This is consistent with the expected global error of $\mathcal{O}(\delta \tau^{2})$ corresponding to an error of $\mathcal{O}(\delta \tau^{3})$ over each time step of the second-order splitting scheme. 
 In Fig.~\ref{wp} we show the propagation of this  wave packet through the curved spacetime.

\begin{figure}[t]
\centering
\includegraphics[width=0.4\textwidth]{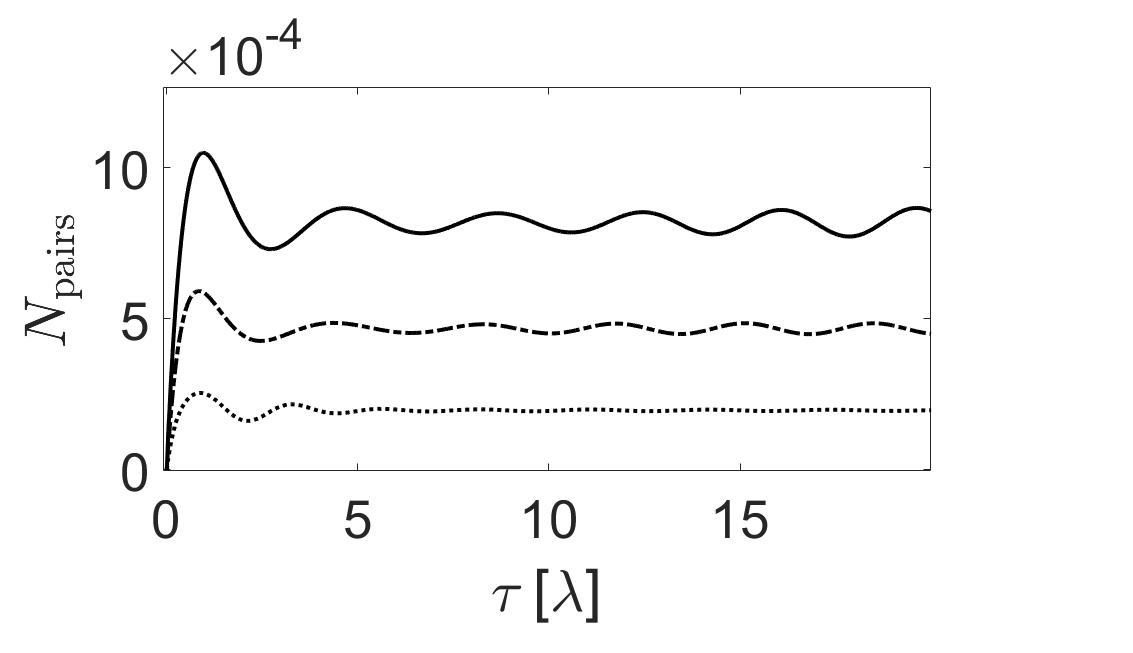}

\caption{Number growth of created fermion-antifermion pairs for different widths of the Gaussian curvature Eq.~\eqref{defo}. For wider curvatures, the production reaches saturation at later times and for greater numbers of created pairs. The dotted line corresponds to $r_0=1 \lambda$, the dash-dotted line corresponds to $r_0=\sqrt{2} \lambda$ and the solid line corresponds to $r_0=\sqrt{3} \lambda$.  The numbers are calculated on a lattice of width $100 \lambda$ and of $N = 2^{11}$ sites.}
\label{nbs_rs}
\end{figure}

\begin{figure}[t]
\centering
\includegraphics[width=0.4\textwidth]{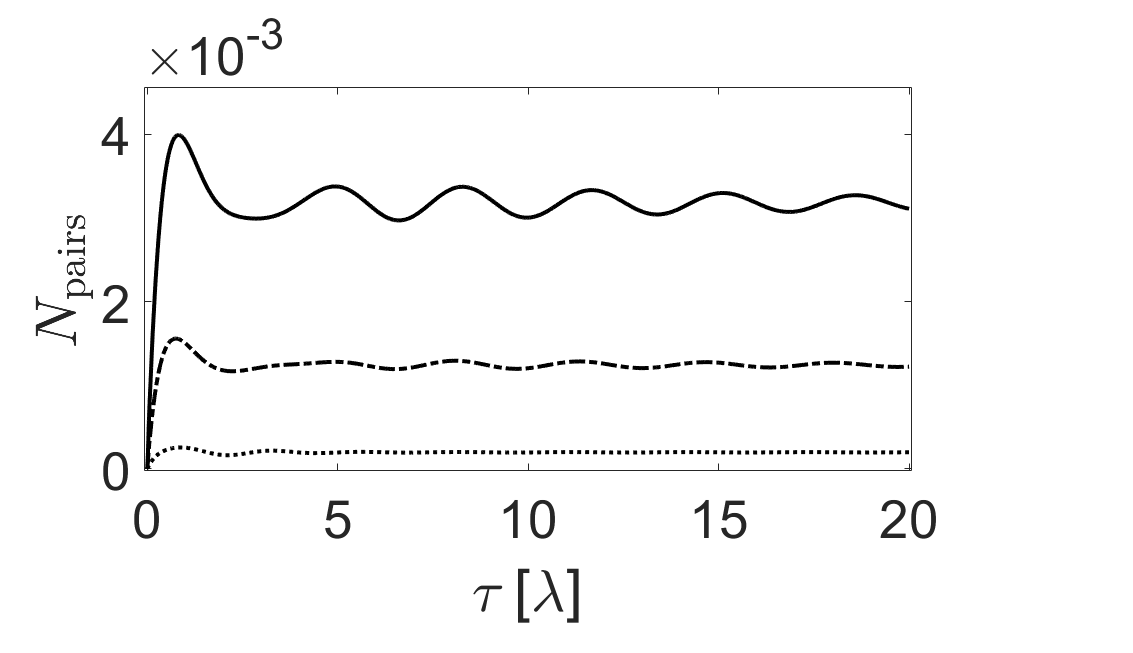}
\caption{Number growth of created fermion-antifermion pairs for different strengths of the Gaussian curvature with $r_0=1 \lambda$. For stronger curvatures, more pairs are created and saturation is reached for greater numbers. The dotted line corresponds to $\beta=0.5$, the dash-dotted line corresponds to $\beta = 0.7$, the solid line corresponds to $\beta = 0.8$ and the dotted line corresponds to $\beta = 0.9$.}
\label{nbs_sts}
\end{figure}

\begin{figure}[b]
\centering
\includegraphics[width=0.4\textwidth]{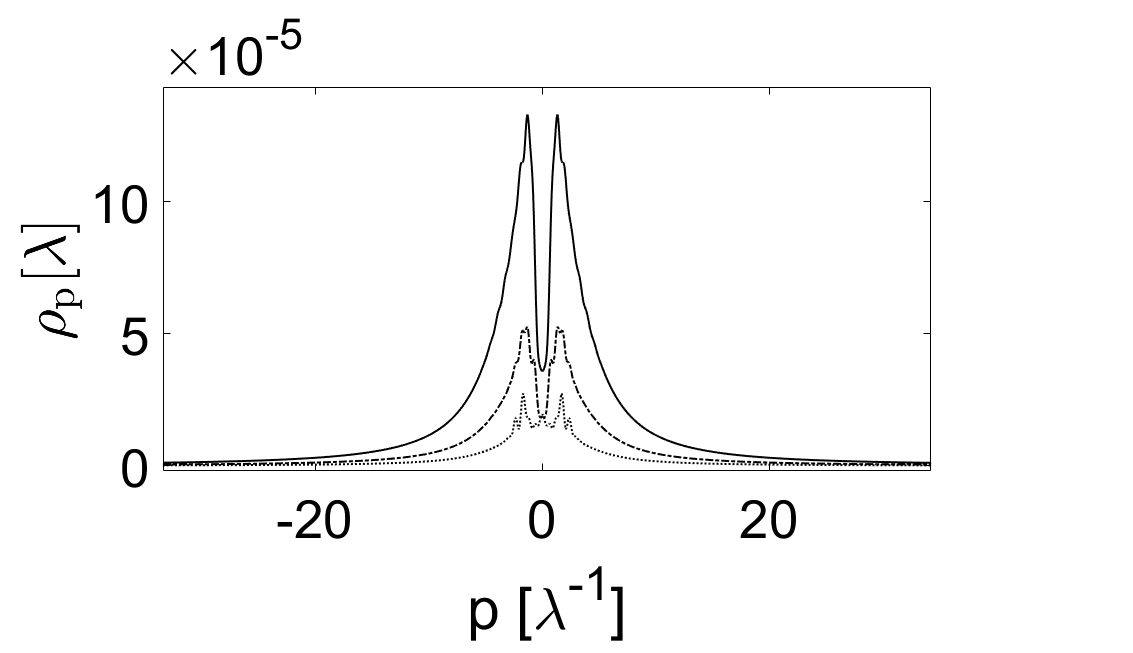}
\caption{Spectrum of created fermion-antifermion pairs calculated at $\tau = 10 \lambda$ for different strengths of the Gaussian curvature with $r_0=1 \lambda$. For stronger curvatures, more modes are created and saturation is reached for greater numbers. The solid line corresponds to $\beta=0.7$, the dash-dotted line corresponds to $\beta=0.6$, and the dotted line corresponds to $\beta = 0.5$.}
\label{rhop_sts}
\end{figure}

In Fig.~\ref{nbs_rs} we plot the total number of created fermion--antifermion
pairs as a function of time. After an initial growth, the production
reaches saturation due to the Pauli blockade. The number of pairs created at which saturation
sets in depends on the spatial width of the curvature deformation; wider curvature profiles lead to greater asymptotic
pair numbers, as the pairs can be excited over a larger spatial volume. Similarly, stronger curvature deformations correspond to larger values of the parameter $\beta$ in Eq.~\eqref{defo} as it increases the value of the Ricci scalar Eq.~(\ref{ricci}). As can be seen in Fig.~\ref{nbs_sts} which shows the number of created pairs as a function of time for different values of $\beta$, increasing $\beta$ results in a
faster growth of the particle number and a higher saturation value. Fig.~\ref{rhop_sts} shows the momentum spectrum of created pairs obtained at late times for different values of $\beta$ too. Greater densities are obtained for greater values of $\beta$.
Physically, stronger curvature deformations couple to a larger number
of fermionic modes, enabling pair production channels that are not
accessible for smaller $\beta$ -- note, however, that this does not delay the onset of Pauli blocking.

\begin{figure}
\centering
\includegraphics[width=0.4\textwidth]{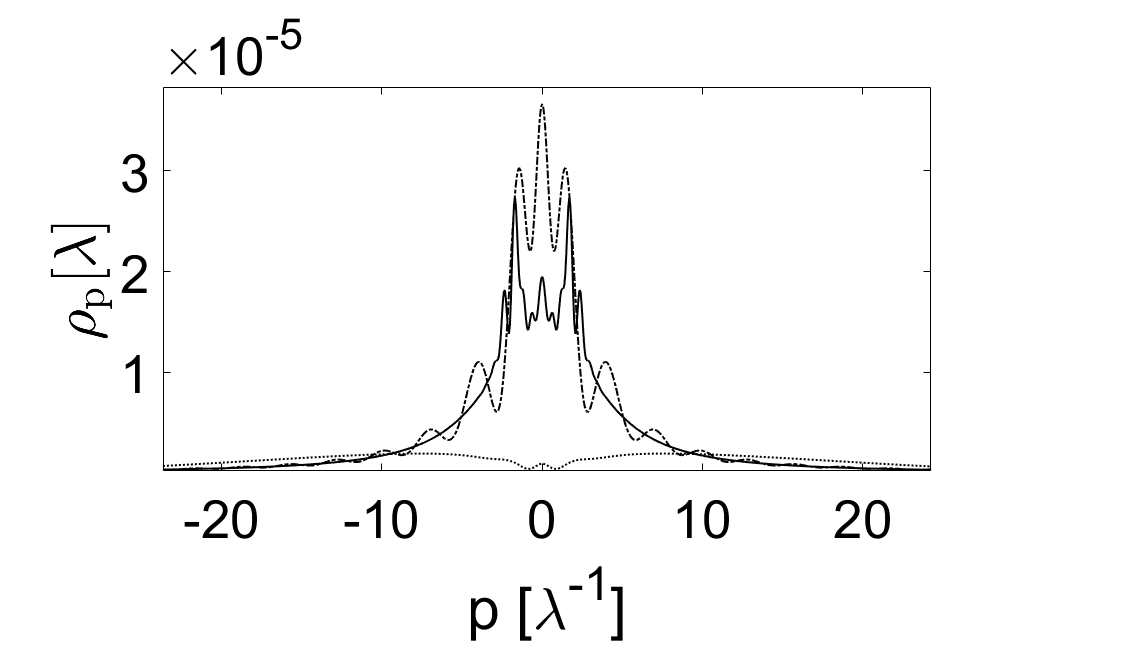}
\includegraphics[width=0.4\textwidth]{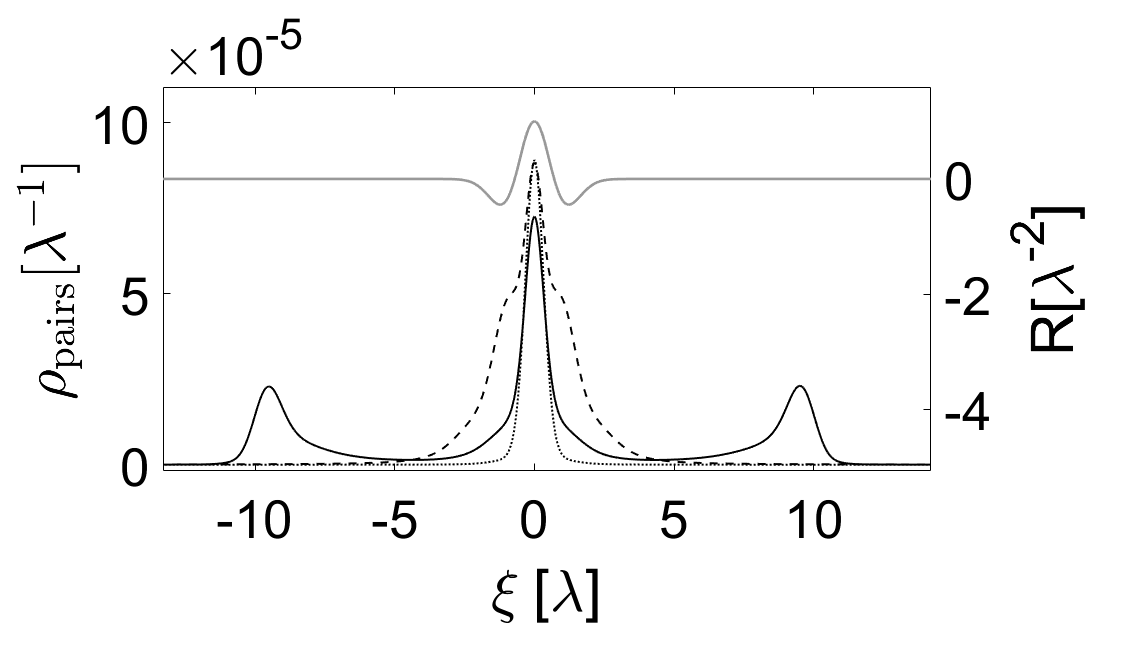}
\caption{ momentum spectrum of created fermion--antifermion pairs (upper panel) and their  number density(lower panel) at three times:
$\tau=0.1\,\lambda$ (dotted line), $\tau=1.3\,\lambda$ (dashed line), and
$\tau=10\,\lambda$ (solid black line), where $\lambda$ is the electron Compton
wavelength. The Ricci scalar $R(\xi)$, Eq.~\eqref{ricci} corresponding to the spacetime curvature given by
Eqs.~\eqref{metricschwa} and~\eqref{defo} with $\beta=\tfrac{1}{2}$ is shown in the upper panel by
the solid gray line. The calculated number densities are physical at late times and far from the spatial variations of the Ricci scalar. Calculations use $N=2^{11}$ lattice sites and a time step
$\delta\tau=0.01\,\lambda$.}
\label{rhosfig}
\end{figure}

In Fig.~\ref{rhosfig} we show the time evolution of the momentum spectrun of the created pairs and of their number density as a function of spatial position, by plotting these quantities for three illustrative times.

The results show that pair creation is initially supported close to the central, positive peak in the Ricci curvature. Pairs later start to form in regions of negative curvature and propagate outwards. Before saturation, the total number of pairs is monotonically increasing (see also Fig. \ref{nbs_conv}). The part of the density created in the regions of negative curvature gradually propagates far from the center of the curvature bump where the calculated number densities become physical, while the part created in the positive region is confined inside the curvature region, blocking more creation at later times (the Pauli blockade). The momentum spectrum shows little temporal variation of the range of momenta of created pairs, but they become more concentrated about the turning points of the scalar curvature momentum as time increases, which tends to reduce the range of momenta carried by the pairs produced at later times.

\section{Discussion}

We have shown that the CQFT formalism can be extended to curved
spacetimes and used to investigate vacuum excitation and real-time
fermion--antifermion pair creation. The particle numbers reported in this
work are defined with respect to the free Dirac basis; they correspond to the excitations that would be observed by a Minkowski observer if the
curvature were removed instantaneously. As is well known in quantum
field theory in curved spacetime \cite{birrell_davies,parker_toms},
particle number is not an invariant quantity and depends on the
particular mode decomposition used to define the vacuum. The goal of the
present work is therefore not to provide a physical particle spectrum in
a specific observational scenario, but rather to establish the
mathematical and numerical framework required for applying CQFT to
curved backgrounds and obtain unambiguous pair creation numbers as seen by inertial observers in asymptotically flat Minkowksi space.

Another physically meaningful particle definition can be obtained by
projecting onto the instantaneous eigenstates of the Hamiltonian or of
the charge-density operator; this will be explored in future work.
Importantly, while the number densities of separate positive- and negative-energy states
depend on the chosen basis, the charge density, which is equal to the difference between
them, is basis-independent in the sense that changing the basis corresponds to changing the vacuum with respect to which we calculate the numbers. This becomes especially relevant when
spatially dependent electromagnetic fields are included, since they can
separate the created charges and give rise to basis-independent quantities such as charge densities and currents. The same CQFT approach can be used to investigate the time evolution of scalar fields in curved spacetimes. This will be investigated in future works. In addition, it would be interesting to employ the CQFT framework to investigate whether universal scaling behavior analogous to that observed in spatially inhomogeneous electromagnetic fields \cite{gies2016critical,gies2017critical} arises near critical points in pair production induced by varying geometries.

Future extensions of the method will therefore include, in addition to the
curved background, spatially varying electromagnetic fields, as well
as implementing projection schemes based on instantaneous
diagonalization of the Hamiltonian. Additionally, numerical improvements
will be implemented to reduce time ordering errors in the time evolution operator
 beyond the time-step refinement used in the present work \cite{grobe_timeorder}.
These extensions will enable the study of more realistic physical
scenarios in which the notion of particle number becomes unambiguous
and directly observable. Moreover, the field self-interaction can be introduced through effective actions taking into consideration the radiative corrections. 

Extending the present CQFT approach to higher spacetime dimensions
faces the fact that, in contrast to the two-dimensional case, generic
spacetimes are not locally conformally flat. Nevertheless, many
physically relevant geometries, including FLRW,
de Sitter, and anti--de Sitter spacetimes, are conformally flat. In such
cases, rescalings analogous to that introduced in
Eq.~\eqref{rescaling} can be employed to obtain a Hermitian Hamiltonian
and a unitary time-evolution operator. By contrast, for spacetimes that
are not conformally flat, such as the Schwarzschild geometry, alternative
strategies must be adopted, perhaps on a case by case basis, in order to render the Dirac equation in a suitable Schrödinger form so that the first-quantized approach can be employed.

\begin{acknowledgments}
M.~Alkhateeb acknowledges funding from the European Union’s Horizon 2020
research and innovation programme under the Marie Skłodowska-Curie
grant agreement No.~101034383. He also thanks the University of Plymouth's hospitality where part of this work was conducted.
Y.~Caudano is a research associate of the Fund for Scientific Research F.R.S--FNRS. The authors are grateful to the anonymous reviewer for comments and suggestions that improved this manuscript.
\end{acknowledgments}

\bibliographystyle{unsrt}
\bibliography{bibliography.bib}


\appendix

\section{Dirac equation in curved spacetime}
\label{App:Curved}
The formulation of the Dirac equation in curved spacetime relies on \J{a} smooth curved spacetime having a Minkowskian tangent space at any point. At each spacetime
point one may introduce a local inertial frame in which the metric takes
the Minkowski form and local Lorentz symmetry is manifest. The relation
between quantities defined in this local Lorentz frame and those defined
in an arbitrary curved coordinate system is provided by the tetrad (or
vielbein) formalism.

In this appendix we review the construction of a covariant derivative
acting on spinor fields in curved spacetime and outline the derivation of the Dirac
equation in a generally covariant form. The strategy is to determine the
spinor connection by requiring that spinor bilinears transform correctly as tensors
under parallel transport. In particular, we will demand that the
probability current constructed from a spinor field transforms as a
vector under local Lorentz transformations and parallel transport.

Throughout this section, Greek indices $(\mu,\nu,\ldots)$ denote curved
spacetime coordinate indices, while Latin indices $(a,b,\ldots)$ denote
indices associated with the local Lorentz (orthonormal) frame.

\subsection{Tetrads and local inertial frames}
\label{tetrads}
At each spacetime point $x$, one may introduce a local inertial (orthonormal)
frame in accordance with the equivalence principle. The tetrad fields
$e^\mu{}_a(x)$ and their inverses $e^a{}_\mu(x)$ relate tensor components in
this local Lorentz frame to those in an arbitrary curved coordinate system.
At a given spacetime point, one may choose local inertial coordinates $y^a$
such that the tetrads coincide with the Jacobian matrices
\begin{equation}
	e^\mu{}_a(x)
	=
	\left.\frac{\partial x^\mu}{\partial y^a}\right|_x,
	\qquad
	e^a{}_\mu(x)
	=
	\left.\frac{\partial y^a}{\partial x^\mu}\right|_x.
\end{equation}
This identification is purely pointwise and does not imply the existence of a
global coordinate transformation; in a generic curved spacetime the tetrad
fields are not globally integrable as coordinate derivatives.
For a vector $j$, the components in the two frames are related by
\begin{equation}
\label{trans_vec}
	j^\mu(x) = e^\mu{}_a(x)\, j^a(x),
	\qquad
	j^a(x) = e^a{}_\mu(x)\, j^\mu(x).
\end{equation}
The tetrads satisfy the orthonormality relations
\begin{equation}
\begin{split}
g_{\mu\nu}(x)\, e^\mu{}_a(x)\, e^\nu{}_b(x) &= \eta_{ab},\\
\eta_{ab}\, e^a{}_\mu(x)\, e^b{}_\nu(x) &= g_{\mu\nu}(x),\\
e^\mu{}_a(x)\, e^a{}_\nu(x) &= \delta^\mu{}_\nu,\\
e^a{}_\mu(x)\, e^\mu{}_b(x) &= \delta^a{}_b.
\end{split}
\end{equation}

where $\eta_{ab}=\mathrm{diag}(1,-1,-1,-1)$ is the Minkowski metric in the
local Lorentz frame.

\subsection{Levi--Civita connection and vector parallel transport}
\label{Gammas}
We assume that spacetime is equipped with the Levi--Civita connection,
which is torsion-free,
\[
\Gamma^\rho{}_{\mu\nu}=\Gamma^\rho{}_{\nu\mu},
\]
and metric compatible,
\[
\nabla_\rho g_{\mu\nu}=0.
\]
The corresponding Christoffel symbols are therefore given by
\begin{equation}
\Gamma^{\rho}{}_{\mu\nu}
=
\frac{1}{2} g^{\rho\sigma}
\left(
\partial_\mu g_{\nu\sigma}
+ \partial_\nu g_{\mu\sigma}
- \partial_\sigma g_{\mu\nu}
\right).
\end{equation}
The parallel transport of a vector expressed in the coordinate basis
takes the form
\begin{equation}
\label{cov_diff}
	j^\mu(x+dx)
	=
	j^\mu(x)
	-
	\Gamma^\mu{}_{\nu\sigma}(x)\, j^\nu(x)\, dx^\sigma.
\end{equation}
It is often convenient to express vector components in the local Lorentz
frame while keeping the spacetime displacement $dx^\mu$ in curved
coordinates. We therefore introduce a ``mixed'' parallel transport law,
\begin{equation}
\label{cov_mix}
	j^a(x+dx)
	=
	j^a(x)
	-
	\omega^a{}_{b\sigma}(x)\, j^b(x)\, dx^\sigma,
\end{equation}
where $\omega^a{}_{b\sigma}$ is the spin connection associated with the
tetrad. The corresponding covariant derivative is
\begin{equation}
	\nabla_\sigma j^a
	=
	\partial_\sigma j^a
	+
	\omega^a{}_{b\sigma}\, j^b.
\end{equation}
The purpose of introducing this mixed covariant derivative is to relate
the parallel transport of vectors in curved spacetime to their transport
in the local Lorentz frame. By requiring consistency between these two
descriptions, we determine the spin connection
$\omega^{a}{}_{b\nu}$ in terms of the tetrads and the Christoffel symbols.

\subsection{Spin connection from tetrads}
\label{omegas}
Using the transformation rule \eqref{trans_vec}, the mixed transport law
\eqref{cov_mix} can be rewritten in terms of curved indices as
\begin{equation}
\label{findspindconnec}
\begin{split}
	j^\mu(x+dx)
	&=
	e^\mu{}_a(x+dx)\, j^a(x+dx) \\
	&=
	j^\mu(x)
	+
	\partial_\nu e^\mu{}_a(x)\, j^a(x)\, dx^\nu \\
	&- 
	e^\mu{}_a(x)\, \omega^a{}_{b\nu}(x)\, j^b(x)\, dx^\nu,
\end{split}
\end{equation}
where we have used
\begin{equation}
	e^\mu{}_a(x+dx)
	=
	e^\mu{}_a(x)
	+
	\partial_\nu e^\mu{}_a(x)\, dx^\nu.
\end{equation}
Comparing Eq.~\eqref{findspindconnec} with the standard coordinate-basis
parallel transport \eqref{cov_diff} and eliminating $j^a$, we obtain
\begin{equation}
	-\Gamma^\mu{}_{\nu\sigma}\, e^\sigma{}_b
	=
	\partial_\nu e^\mu{}_b
	-
	e^\mu{}_a\, \omega^a{}_{b\nu}.
\end{equation}
Solving for the spin connection yields
\begin{equation}
	\omega^a{}_{b\nu}
	=
	e^a{}_\mu\, \partial_\nu e^\mu{}_b
	+
	e^a{}_\mu\, e^\sigma{}_b\, \Gamma^\mu{}_{\nu\sigma}.
\end{equation}
This is the standard expression for the spin connection in terms of the
tetrads and the Levi--Civita connection.

\subsection{Antisymmetry of the spin connection}
\label{omegaantisy}
The spin connection is antisymmetric in its local Lorentz indices. This
property follows from metric compatibility in the local frame, or
equivalently from the requirement that parallel transport preserves the
local Lorentz scalar product.

To demonstrate this explicitly, consider the squared length of a vector
after parallel transport:
\begin{equation}
\begin{split}
	j^a(x+dx)\, \eta_{ab}\, j^b(x+dx)
	&=
	\eta_{ab} j^a(x) j^b(x)
	\\
    & -
	\eta_{ab} j^a \omega^b{}_{d\nu} j^d\, dx^\nu
		\\
    &
    -
	\eta_{ab} j^b \omega^a{}_{c\nu} j^c\, dx^\nu.
\end{split}
\end{equation}
Since parallel transport preserves lengths, the $\mathcal{O}(dx)$ terms
must cancel, implying
\begin{equation}
	\eta_{ab} j^a \omega^b{}_{d\nu} j^d
	=
	-
	\eta_{ab} j^b \omega^a{}_{c\nu} j^c.
\end{equation}
Lowering the first Lorentz index of the spin connection,
$\omega_{ad\nu} \equiv \eta_{ab} \omega^b{}_{d\nu}$, this becomes
\begin{equation}
	j^a \omega_{ad\nu} j^d
	=
	-
	j^a \omega_{da\nu} j^d.
\end{equation}
Because $j^a j^d$ is symmetric under $a \leftrightarrow d$, this relation
can hold for arbitrary $j^a$ only if
\begin{equation}
	\omega_{ad\nu} = - \omega_{da\nu}.
\end{equation}

\subsection{Spinor parallel transport and spinor connection}
\label{Omegas}
We now turn to spinor fields. We assume that the parallel transport of a
spinor $\psi$ is governed by a matrix-valued connection $\Omega_\nu$,
\begin{equation}
	\psi(x+dx)
	=
	\psi(x)
	-
	\Omega_\nu(x)\, \psi(x)\, dx^\nu.
\end{equation}
The adjoint spinor $\bar\psi=\psi^\dagger \gamma^0$ transforms as
\begin{equation}
	\bar\psi(x+dx)
	=
	\bar\psi(x)
	-
	\psi^\dagger(x)\, \Omega_\nu^\dagger(x)\, \gamma^0\, dx^\nu\,,
\end{equation}
while the scalar bilinear
\begin{equation}
	S(x)=\bar\psi(x)\psi(x)
\end{equation}
must transform as a scalar under parallel transport. Requiring
$\nabla_\nu S=\partial_\nu S$ implies
\begin{equation}
	\Omega_\nu
	=
	-
	\gamma^0 \Omega_\nu^\dagger \gamma^0.
\end{equation}

Next, consider the probability current in the local Lorentz frame,
\begin{equation}
	j^a(x)
	=
	\bar\psi(x)\gamma^a\psi(x),
\end{equation}
which must transform as a Lorentz vector. Computing its parallel
transport using the spinor connection gives
\begin{equation}
	j^a(x+dx)
	=
	j^a(x)
	+
	\bar\psi(x)\,[\Omega_\nu,\gamma^a]\,\psi(x)\, dx^\nu.
\end{equation}
On the other hand, vector parallel transport requires
\begin{equation}
	j^a(x+dx)
	=
	j^a(x)
	-
	\omega^a{}_{b\nu}\, j^b(x)\, dx^\nu.
\end{equation}
Equating these expressions yields the well-known condition
\begin{equation}
	[\gamma^a,\Omega_\nu]
	=
	\omega^a{}_{b\nu}\, \gamma^b.
\end{equation}
A solution consistent with this relation is
\begin{equation}
\label{Om}
	\Omega_\nu
	=
	-\frac{i}{4}\, \omega_{ab\nu}\, \sigma^{ab},
\end{equation}
where
\begin{equation}
	\sigma^{ab}
	=
	\frac{i}{2}[\gamma^a,\gamma^b].
\end{equation}
This expression satisfies the required commutation relation, as may be
verified using the gamma-matrix algebra.

\subsection{Dirac equation in curved spacetime}
\label{diraceqapp}
The curved-spacetime gamma matrices are defined by
\[
\gamma^\mu(x) \equiv e^\mu{}_a(x)\,\gamma^a,
\]
which ensures that they satisfy the spacetime Clifford algebra
\(\{\gamma^\mu,\gamma^\nu\}=2g^{\mu\nu}\mathbb{I}\) and that the Dirac
current transforms as a spacetime vector, and the spinor covariant derivative is defined by
\begin{equation}
	\nabla_\nu \psi
	=
	\partial_\nu \psi
	+
	\Omega_\nu \psi.
\end{equation}
The Dirac equation in curved spacetime then takes the generally covariant
form
\begin{equation}
	i \gamma^\nu \nabla_\nu \psi
	-
	m \psi
	=
	0,
\end{equation}
or explicitly,
\begin{equation}
	i e^\nu{}_a \gamma^a
	\left(
	\partial_\nu + \Omega_\nu
	\right)\psi
	-
	m\psi
	=
	0.
\end{equation}
with $\Omega_\nu$ given by Eq.~\eqref{Om}.

\section{Local conformal flatness of \(1\!+\!1\)-dimensional metrics}
\label{app:conformal_flatness}

In this appendix we show that any \(1+1\)-dimensional Lorentzian metric of the form
\begin{equation}
\label{eq:original_metric}
ds^2
=
\alpha_1(\tau,\xi)\,d\tau^2
-
\alpha_2(\tau,\xi)\,d\xi^2,
\qquad
\alpha_1 \ge 0,\quad \alpha_2 \ge 0,
\end{equation}
is locally conformally flat. That is, in a suitable local coordinate system
\((\tau',\xi')\), the metric can be written as
\begin{equation}
\label{eq:conformal_metric}
ds^2
=
\alpha(\tau',\xi')\,
\bigl(d\tau'^2-d\xi'^2\bigr),
\end{equation}
with a positive conformal factor \(\alpha\).

Throughout this appendix we work locally on an open set where
\(\alpha_1(\tau,\xi)>0\) and \(\alpha_2(\tau,\xi)>0\), so that the metric has
Lorentzian signature.

\subsection{Null decomposition}

Define the function
\begin{equation}
\label{eq:c_def}
c(\tau,\xi)
:=
\sqrt{\frac{\alpha_2(\tau,\xi)}{\alpha_1(\tau,\xi)}}.
\end{equation}
Then the metric \eqref{eq:original_metric} can be factorized as
\begin{equation}
\label{metric_factorized}
\begin{split}
ds^2
=
&\alpha_1(\tau,\xi)
\bigl(d\tau^2-c(\tau,\xi)^2\,d\xi^2\bigr)
\\
=&\alpha_1(\tau,\xi)\,
\bigl(d\tau-c\,d\xi\bigr)
\bigl(d\tau+c\,d\xi\bigr).
\end{split}
\end{equation}

Introduce the two null 1-forms
\begin{equation}
\label{eq:null_forms}
\omega_- := d\tau-c\,d\xi,
\qquad
\omega_+ := d\tau+c\,d\xi,
\end{equation}
so that
\begin{equation}
\label{eq:metric_null}
ds^2=\alpha_1\,\omega_-\,\omega_+ .
\end{equation}
It remains to seek local coordinates adapted to these null directions.
\subsection{Integrating factors and null coordinates}
We look for nonvanishing functions
\(\mu(\tau,\xi)\) and \(\nu(\tau,\xi)\) such that the 1-forms
\begin{equation}
\label{eq:def_uv}
du := \mu\,\omega_- = \mu\,(d\tau-c\,d\xi),
\qquad
dv := \nu\,\omega_+ = \nu\,(d\tau+c\,d\xi),
\end{equation}
are exact.

The condition that \(du\) and \(dv\) be closed,
\(d(du)=0\) and \(d(dv)=0\), yields the first-order linear partial
differential equations
\begin{align}
\label{eq:mu_eq}
\partial_\xi \mu + \partial_\tau(\mu c) &= 0,\\[4pt]
\label{eq:nu_eq}
\partial_\xi \nu - \partial_\tau(\nu c) &= 0.
\end{align}
These equations are transport equations along the characteristic curves
defined by
\begin{equation}
\label{eq:characteristics}
d\tau = \pm c(\tau,\xi)\,d\xi,
\end{equation}
which coincide with the null curves of the metric.
Standard local existence results for first-order linear PDEs ensure that
smooth, nonvanishing solutions \(\mu\) and \(\nu\) exist in a sufficiently
small neighborhood, given appropriate initial data.
Consequently, the functions \(u(\tau,\xi)\) and \(v(\tau,\xi)\) define
a local coordinate system.

From \eqref{eq:def_uv} we obtain
\begin{equation}
\label{eq:inverse_forms}
d\tau-c\,d\xi=\frac{1}{\mu}\,du,
\qquad
d\tau+c\,d\xi=\frac{1}{\nu}\,dv.
\end{equation}
Substituting into \eqref{eq:metric_null} gives
\begin{equation}
\label{eq:metric_uv}
ds^2
=
\frac{\alpha_1(\tau,\xi)}{\mu(\tau,\xi)\,\nu(\tau,\xi)}\,du\,dv
=
\theta^2(u,v)\,du\,dv,
\end{equation}
where
\begin{equation}
\label{eq:Omega_def}
\theta^2(u,v)
:=
\frac{\alpha_1(\tau(u,v),\xi(u,v))}
{\mu(\tau,\xi)\,\nu(\tau,\xi)} .
\end{equation}
\smallskip

\subsection{Conformally flat coordinates}

Finally, define new coordinates \((\tau',\xi')\) by
\begin{equation}
\label{eq:tauprime_xiprime}
\tau'=\frac{v+u}{2},
\qquad
\xi'=\frac{v-u}{2}.
\end{equation}
A direct computation shows that
\begin{equation}
\label{eq:du_dv}
du\,dv = d\tau'^2-d\xi'^2.
\end{equation}
Therefore, the metric \eqref{eq:metric_uv} becomes
\begin{equation}
\label{eq:final_metric}
ds^2
=
\theta^2(u,v)\,
\bigl(d\tau'^2-d\xi'^2\bigr).
\end{equation}
Defining
\begin{equation}
\label{eq:theta_def}
\alpha(\tau',\xi')
:=
\theta^2\bigl(u(\tau',\xi'),v(\tau',\xi')\bigr),
\end{equation}
we arrive at the conformally flat form
\begin{equation}
\label{eq:conformal_final}
ds^2
=
\alpha(\tau',\xi')\,
\bigl(d\tau'^2-d\xi'^2\bigr).
\end{equation}

This completes the proof of the well-known fact that any \(1+1\)-dimensional Lorentzian metric of
the form \eqref{eq:original_metric} is locally conformally flat.

\end{document}